\documentclass[10pt,aps,twocolumn,prc,nofootinbib,noshowkeys,longbibliography,superscriptaddress,floatfix]{revtex4-1}
\usepackage{amsmath,amsfonts,amsthm,amssymb,physics,graphicx}
\usepackage[usenames,dvipsnames]{xcolor}
\usepackage[colorlinks=true,
			linktocpage=true,
			linkcolor=blue,
			citecolor=red,
			urlcolor=blue]{hyperref}
\usepackage{yfonts}

\def\p{{\cal P}}
\def\e{{\cal E}}
\def\taur{\tau_{\rm R}}

\begin{document}

\preprint{}

\title{Relativistic second-order viscous hydrodynamics from kinetic theory with extended relaxation-time approximation}

\author{Dipika Dash}
\email{dipika.dash@niser.ac.in}
\affiliation{School of Physical Sciences, National Institute of Science Education and Research, An OCC of Homi Bhabha National Institute, Jatni 752050, India}
\author{Sunil Jaiswal}
\email{jaiswal.61@osu.edu}
\affiliation{Department of Nuclear and Atomic Physics, Tata Institute of Fundamental Research, Mumbai 400005, India}
\affiliation{Department of Physics, The Ohio State University, Columbus, Ohio 43210-1117, USA}
\author{Samapan Bhadury}
\email{samapan.bhadury@uj.edu.pl}
\affiliation{Institute of Theoretical Physics, Jagiellonian University, ul. St. \L ojasiewicza 11, 30-348 Krakow, Poland}
\author{Amaresh Jaiswal}
\email{a.jaiswal@niser.ac.in}
\affiliation{School of Physical Sciences, National Institute of Science Education and Research, An OCC of Homi Bhabha National Institute, Jatni 752050, India}

%-----------------------------------------------------------------------

\date{\today}

\begin{abstract} 
We use the extended relaxation time approximation for the collision kernel, which incorporates a particle-energy dependent relaxation time, to derive second-order viscous hydrodynamics from the Boltzmann equation for a system of massless particles. The resulting transport coefficients are found to be sensitive to the energy dependence of the relaxation time and have significant influence on the fluid's evolution. Using the derived hydrodynamic equations, we study the evolution of a fluid undergoing (0+1)-dimensional expansion with Bjorken symmetry and investigate the fixed point structure inherent in the equations. Further, by employing a power law parametrization to describe the energy dependence of the relaxation time, we successfully reproduce the stable free-streaming fixed point for a specific power of the energy dependence. The impact of the energy-dependent relaxation time on the processes of isotropization and thermalization of an expanding plasma is discussed.
\end{abstract}

%\pacs{25.75.-q, 24.10.Nz, 47.75+f}
% 25.75.-q Relativistic heavy-ion collisions
% 24.10.Nz Hydrodynamic models
% 47.75.+f Relativistic fluid dynamics

\maketitle
%-----------------------------------------------------------------------

%=============================================================
\section{Introduction}
\vspace{-.2cm}
%=============================================================

The relativistic Boltzmann equation is a transport equation that governs the space-time evolution of the single-particle phase-space distribution function. It is capable of accurately describing the collective dynamics of the system in the limit of small mean free path and therefore has been employed extensively to formulate the theory of relativistic hydrodynamics~\cite{Muronga:2006zx, York:2008rr, Betz:2008me, Romatschke:2009im, Denicol:2012cn, Florkowski:2013lya, Weickgenannt:2020aaf}. However, solving the Boltzmann equation directly is challenging due to the complicated integro-differential nature of the collision term, which involves the integral of the product of distribution functions. Over several decades, various approximations have been proposed to simplify the collision term in the linearized regime. In 1969, following earlier works by Bhatnagar-Gross-Krook~\cite{Bhatnagar:1954zz} and Welander~\cite{Welander}, Marle introduced a relaxation time approximation for non-relativistic systems~\cite{marle1969}. However, Marle's version was not applicable to massless particles and was ill-defined in the relativistic limit. Anderson and Witting resolved these issues by generalizing Marle's model to the relativistic regime, qualitatively recovering the results obtained using Grad's method of moments in the relativistic limit~\cite{anderson1974relativistic}. These models, introduced by Marle and Anderson-Witting, incorporate a collision time scale known as the relaxation time. The Anderson-Witting model requires the relaxation time to be independent of particle momenta, making it straightforward to apply in the formulation of relativistic dissipative hydrodynamics.

The Anderson-Witting model achieves enormous simplification by approximating that collisions drive the system towards local equilibrium exponentially without explicitly describing the interaction mechanism of the microscopic constituents. This approximation provides a highly accurate description of the collective dynamics for systems close to equilibrium. In the following, we will refer to the Anderson-Witting model as the relaxation-time approximation (RTA). Despite its simplistic nature, RTA and its variations have proven to be immensely useful and have been extensively employed in formulating relativistic dissipative hydrodynamics as well as in deriving transport coefficients~\cite{Denicol:2010xn, Jaiswal:2013npa, Jaiswal:2014isa, Mohanty:2018eja, Kurian:2020qjr, Gabbana:2017uvc, Bhadury:2020ngq, Panda:2020zhr, Vyas:2022hkm, Singha:2023eia}. Recently, it has also been applied to study the domain of applicability of hydrodynamics~\cite{Heller:2016rtz, Blaizot:2017ucy, Florkowski:2017olj, Strickland:2018ayk, Chattopadhyay:2018apf, Kurkela:2019set, Blaizot:2019scw, Blaizot:2020gql, Dash:2020zqx, Jaiswal:2020hvk, Soloviev:2021lhs, Blaizot:2021cdv, Chattopadhyay:2021ive, Jaiswal:2021uvv, Jaiswal:2022udf, Chattopadhyay:2022sxk, Ambrus:2022qya, Jankowski:2023fdz}. This simple model appears to capture effective microscopic interactions across a wide range of theories.

When deriving dissipative hydrodynamic equations from kinetic theory using the RTA approximation, it is typically assumed that the relaxation time is independent of particle energy (or momentum). Additionally, one is constrained to work in the Landau frame to ensure the preservation of macroscopic conservation laws. However, in realistic systems, the collision time scale generally depends on the microscopic interactions~\cite{Dusling:2009df, Chakraborty:2010fr, Dusling:2011fd, Kurkela:2017xis}. Introducing an energy-dependent relaxation time leads to a violation of microscopic conservation laws in the Landau frame. As a result, there has been considerable interest in developing a consistent formulation of relativistic dissipative hydrodynamics with an energy-dependent relaxation-time approximation for the Boltzmann equation that satisfies both microscopic and macroscopic conservation laws~\cite{Teaney:2013gca, Mitra:2020gdk, Rocha:2021zcw, Dash:2021ibx}.

Relativistic viscous hydrodynamics-based multistage dynamical models have demonstrated success in accurately describing a broad spectrum of soft hadronic observables in heavy-ion collisions~\cite{Gale:2013da, DerradideSouza:2015kpt, Schenke:2020mbo}. The hydrodynamics stage of the evolution encompasses the deconfined quarks and gluons regime at high temperatures, the phase transition, and the hadron gas phase~\cite{Heinz:2013th}. The dynamical properties of the evolving non-equilibrium nuclear matter are governed by a set of transport coefficients, such as the shear and bulk viscosities~\cite{Romatschke:2007mq, Song:2007fn, Buchel:2008uu, Denicol:2009am, Ryu:2015vwa}. These transport coefficients play a crucial role in explaining the hadronic observables in heavy-ion collisions. Thus, a major goal of heavy-ion phenomenology is to extract the temperature dependence of these transport coefficients for the evolving nuclear matter, and considerable efforts have been made to determine these coefficients from various aspects. Most phenomenological studies adopt parameterized forms for the shear and bulk viscosities~\cite{Akiba:2015jwa, Denicol:2015nhu, Vujanovic:2017psb, Karmakar:2023ity}. Recent studies have employed Bayesian methods to obtain these parameters and have provided bounds on the transport coefficients~\cite{Bernhard:2015hxa, Bernhard:2016tnd, Bernhard:2019bmu, JETSCAPE:2020shq, Nijs:2020ors, JETSCAPE:2020mzn}. However, since these parametrizations do not stem from microscopic considerations, the predictability of such models is limited. Additionally, the second-order transport coefficients utilized in these hydrodynamic models are obtained for specific interactions, and, as a result, they may fail to accurately capture the system's behavior during its evolution. 

In our recent work~\cite{Dash:2021ibx}, we presented a framework for the consistent derivation of relativistic dissipative hydrodynamics from the Boltzmann equation, incorporating a particle energy-dependent relaxation time by extending the Anderson-Witting relaxation-time approximation\footnote{%
    In a recent work \cite{Kamata:2022ola}, the transseries structure of ERTA was explored.}.  
Within this extended RTA (ERTA) framework, we derived the first-order hydrodynamic equations and demonstrated that the hydrodynamic transport coefficients can exhibit significant variations with the energy dependence of the relaxation time. Notably, the ERTA framework allows for the adjustment of interaction characteristics by tuning the energy dependence of the relaxation time, enabling a partial description of the transition from deconfined quark-gluon plasma at high temperatures to a weakly interacting gas of hadrons at lower temperatures. While the formulation presented in Ref.~\cite{Dash:2021ibx} successfully incorporates an energy-dependent relaxation time into the RTA, it still suffers from the well-known issue of acausality in first-order relativistic hydrodynamics within the Landau frame~\cite{Hiscock:1983zz, Hiscock:1985zz, Bemfica:2017wps, Kovtun:2019hdm, Bemfica:2019knx, Hoult:2020eho, Gavassino:2020ubn, Hoult:2021gnb, Biswas:2022cla}. Consequently, there is a need for a second-order theory that addresses this issue \cite{Israel:1976tn, Israel:1979wp}, allowing for its application in heavy-ion collision simulations.

In the present study, we employ the ERTA framework to derive second-order hydrodynamic equations in the Landau frame for a conformal system without conserved charges, incorporating an energy-dependent relaxation time. The second-order transport coefficients are found to be sensitive to the energy dependence of the relaxation time. We focus on a boost-invariant flow in (0+1) dimensions and investigate the fixed point structure of the hydrodynamic equations. Our analysis reveals that the location of the free-streaming fixed points is influenced by the energy dependence of the relaxation time. By employing a power law parametrization to describe this energy dependence, we successfully reproduce the stable free-streaming fixed point for a specific power of the energy dependence. Furthermore, we explore the impact of the energy-dependent relaxation time on the processes of isotropization and thermalization of a boost invariant expanding plasma.

This paper is organized as follows: In Sec.~\ref{sec:overview} we review the basic hydrodynamic equations for a conformal, chargeless fluid. Sec.~\ref{sec:fo_hydro} briefly summarizes the results of Ref.~\cite{Dash:2021ibx} and outlines the steps necessary to derive second-order hydrodynamic equations, which we present in Sec.~\ref{sec:2ndordr}. Appendix~\ref{app:LFC} contains the derivation of the results stated in Sec.~\ref{sec:lfc}. In Appendix~\ref{app:MCO3}, we show that microscopic conservation holds till third order following the prescription outlined in Sec.~\ref{sec:2ndordr}. In Sec.~\ref{sec:bjorken}, we consider Bjorken flow and study the effect of the energy dependence of the relaxation time on systems' thermalization. We summarize our results in Sec.~\ref{sec:summary}.

%========================================================
\vspace{-.4cm}
\section{Overview} \label{sec:overview}
\vspace{-.4cm}
%========================================================

The energy-momentum tensor for a system of massless particles with no net conserved charge can be expressed in terms of the single-particle phase–space distribution function, $f(x,p)$, as
\begin{equation} \label{cons_emt}
T^{\mu\nu} = \int \mathrm{dP}\, p^\mu p^\nu f  = \e u^\mu u^\nu -\p \Delta^{\mu\nu} + \pi^{\mu\nu} ,
\end{equation}
where $\mathrm{dP} = d^3 \vec{p}/\left[(2\pi)^3 E_p\right]$ is the invariant momentum-space integration measure with $E_p$ representing the particle energy which is equal to the magnitude of the particle three-momenta for massless particles, $E_p=|\Vec{p}|$. The projection operator $\Delta^{\mu\nu}=g^{\mu\nu}-u^\mu u^\nu$ is orthogonal to the hydrodynamic four-velocity $u^\mu$ defined in the Landau frame: $u_\nu T^{\mu\nu}=\e u^\mu$, where $\e$ is the energy density. In the above equation, $\mathcal{P}$ is the thermodynamic pressure and $\pi^{\mu\nu}$ is the shear viscous stress. We work with the Minkowskian metric tensor $g^{\mu\nu}\equiv\mathrm{diag}(+,-,-,-)$.

The energy-momentum conservation $\partial_\mu T^{\mu\nu} =0$ yields the fundamental evolution equations for $\e$ and $u^\mu$ as,
\begin{align}
\dot\e + (\e+\p)\theta - \pi^{\mu\nu}\sigma_{\mu\nu} &= 0,  
\label{evol1}\\
(\e+\p)\dot u^\alpha - \nabla^\alpha \p + \Delta^\alpha_\nu \partial_\mu \pi^{\mu\nu}  &= 0, 
\label{evol2}
\end{align}
Here we use the standard notation $\dot A=u^\mu\partial_\mu A$ for co-moving derivatives, $\theta\equiv\partial_\mu u^\mu$ for the expansion scalar, $\sigma^{\mu\nu}\equiv\frac{1}{2}(\nabla^\mu
u^\nu+\nabla^\nu u^\mu)-\frac{1}{3}\theta\Delta^{\mu\nu}$ for the velocity stress tensor, and $\nabla^\alpha=\Delta^{\mu\alpha} \partial_\mu$ for space-like derivatives.

We consider the equilibrium momentum distribution function to have the Maxwell-Boltzmann distribution in the local rest frame of the fluid, $f_{\rm eq}=\exp\left[-(u\cdot p)/T \right]$. The equilibrium energy density then takes the form,
\begin{equation}\label{en_den}
    \e_0 = u_\mu u_\nu \int \mathrm{dP}\, p^\mu p^\nu  f_{\rm eq} = \frac{3T^4}{\pi^2}.
\end{equation}
For an out-of-equilibrium system, the temperature $T$ is an auxiliary quantity which we define using the matching condition $\e \equiv \e_0$.
Also, the thermodynamic pressure and entropy density are given by,
\begin{align}
    \p &= -\frac{1}{3}\Delta_{\mu\nu} \int \mathrm{dP}\, p^\mu p^\nu f_{\rm eq} = \frac{T^4}{\pi^2} \,, 
\label{prs} \\
{\cal S} &= \frac{\e+\p}{T} =\frac{4T^3}{\pi^2} \,. \label{ent_den}
\end{align}
The evolution of temperature is obtained from the hydrodynamic equations of motion~\eqref{evol1} and \eqref{evol2},
\begin{align}
   \dot{\beta} &=\frac{\beta \theta}{3}-\frac{\beta}{3(\e+\p)}\pi^{\mu \nu}\sigma_{\mu \nu}\,,
\label{beta_evol_1}\\
    \nabla^{\mu}\beta &= -\beta \dot{u}^\mu-\frac{\beta}{\e+\p} \Delta^\mu_\alpha \partial_\nu \pi^{\alpha \nu}\,,
\label{beta_evol_2}
\end{align}
where $\beta=1/T$. 

The non-equilibrium phase-space distribution function can be written as $f=f_{\rm eq}+\Delta f$, where $\Delta f$ represents the out-of-equilibrium correction to the distribution function. Using Eq.~\eqref{cons_emt} the shear stress tensor $\pi^{\mu\nu}$ can be expressed in terms of $\Delta f$ as
\begin{align}\label{SST}
\pi^{\mu\nu} &= \Delta^{\mu\nu}_{\alpha\beta} \int \mathrm{dP}\, p^\alpha p^\beta\, \Delta f\,, 
\end{align}
where $\Delta^{\mu\nu}_{\alpha\beta}\equiv \frac{1}{2}\left(\Delta^{\mu}_{\alpha}\Delta^{\nu}_{\beta} + \Delta^{\mu}_{\beta}\Delta^{\nu}_{\alpha}\right) - \frac{1}{3}\Delta^{\mu\nu}\Delta_{\alpha\beta}$ is a doubly symmetric and traceless projection operator orthogonal to $u^\mu$ as well as $\Delta^{\mu\nu}$. The evolution of the shear stress tensor depends on the evolution of the distribution function. In this work, we consider the evolution of the distribution function to be governed by the Boltzmann equation with the collision term, $\textfrak{C}[f]$, in the Extended Relaxation Time Approximation (ERTA) \cite{Teaney:2013gca, Dash:2021ibx},
%sb: Capitalized the T of time and A of approximation.
\begin{align}\label{ERTA}
    p^\mu \partial_\mu f = \textfrak{C}[f] = - \frac{(u\cdot p)}{\taur(x,p)} \left(f - f_{\rm eq}^*\right) \,,
\end{align}
where the relaxation time, $\taur(x,p)$, may depend on the particle momenta. The equilibrium distribution function is considered to be of the Maxwell-Boltzmann form in the `thermodynamic frame', $f_{\rm eq}^* = \exp \left[-\left(u^* \cdot p\right)/T^*\right]$. Here the thermodynamic frame is defined to be the local rest frame of a time-like four vector $u_{\mu}^*$ which need not necessarily correspond to the hydrodynamic four-velocity $u_\mu$, and $T^*$ is the temperature in the local rest frame of $u_{\mu}^*$ (see Ref.~\cite{Dash:2021ibx} for a detailed discussion).

We briefly review the derivation of first-order shear stress from the above kinetic equation in the next section.

%========================================================
\section{First-order hydrodynamics}\label{sec:fo_hydro}
\vspace{-.3cm}
%========================================================

We employ  Chapman-Enskog-like expansion about hydrodynamic equilibrium%
    \footnote{We shall refer to $f_{\rm eq} = \exp\left[-(u\cdot p)/T \right]$ as the hydrodynamic equilibrium distribution function with $u^\mu$ being the fluid four-velocity and $T$ the local fluid temperature in the local rest frame of $u^\mu$.} 
to iteratively solve the ERTA Boltzmann equation \eqref{ERTA},
\begin{align}
    f= f_{\rm eq}+ \delta f_{(1)}+\delta f_{(2)}+\delta f_{(3)}+ \cdots
\end{align}
Here $\delta f_{(i)}$ represents the $i$th order gradient correction to the hydrodynamic equilibrium distribution function. The correction to the distribution function to the first order is 

\vspace{-.6cm}
\begin{equation}\label{deltaf1}
    \delta f_{(1)} = \delta f^*_{(1)} +\frac{\taur}{T} \frac{p^{\mu} p^{\nu}}{u\cdot p} \sigma_{\mu\nu} f_{\rm eq} \,,
\end{equation}
where we have replaced the derivatives of temperature with the derivatives of fluid velocity using Eqs.~(\ref{beta_evol_1},\ref{beta_evol_2}) consistently keeping terms till first order in gradients, and have defined $\delta f^* \equiv f_{\rm eq}^* - f_{\rm eq}$. Defining  $T^* \equiv T+\delta T$ and $u^{\mu}_* \equiv u^\mu + \delta u^\mu $, we obtain the first-order correction $\delta f_{(1)}^*$ by Taylor expanding $f_{\rm eq}^*$ about $u^\mu$ and $T$,
 \begin{align}\label{deltaf1*}
    \delta f^*_{(1)} = \left(- \frac{(\delta u \cdot p)}{T} + \frac{(u\cdot p) \delta T}{T^2} \right)f_{\rm eq} \,.
\end{align}
Using Eqs.~\eqref{deltaf1} and \eqref{deltaf1*}, the quantities $\delta u^\mu$ and $\delta T$ are obtained by imposing the Landau frame conditions, $u_\nu T^{\mu\nu} = \e u^\mu$, and the matching condition, $\e=\e_0$. We find that these quantities vanish for a system of massless and chargeless particles at first-order in gradients, and the resulting first-order correction is given by
\begin{align}\label{Deltaf1}
\delta f_{(1)} = \frac{\tau_{\rm R}}{T} \frac{p^{\mu} p^{\nu}}{u\cdot p} \sigma_{\mu\nu} f_{\rm eq}\,.
\end{align}

It can be easily checked that the microscopic conservation of energy-momentum at first order holds by taking the first momentum-moment of the Boltzmann equation~\eqref{ERTA} with $f\mapsto f_{(1)}=f_{\rm eq} + \delta f_{(1)}$, 
\begin{align}
    \partial_\mu \int \mathrm{dP}\, p^\mu p^\nu  f_{(1)} &= - \int \mathrm{dP}\, \frac{u\cdot p}{\tau_{\rm R}}  p^\nu \left(f_{(1)} - f_{\rm eq}^* \right)
    \nonumber\\
    \implies \partial_\mu T^{\mu\nu}_{(1)} &= - \frac{\sigma_{\alpha\beta}}{T} \int \mathrm{dP}\, p^\nu  p^{\alpha} p^{\beta}  f_{\rm eq} =0\,. \label{del-T^mn_1}
\end{align}

Using $\delta f_{(1)}$ obtained in Eq.~\eqref{Deltaf1}, the expression of shear stress tensor from the definition~\eqref{SST} is obtained to be \cite{Dash:2021ibx},

\vspace{-.6cm}
\begin{equation}\label{shear_NS}
\pi^{\mu\nu} = 2\eta \sigma^{\mu\nu},
\end{equation}
where $\eta = K_{3,2}/T$ is the coefficient of shear viscosity. We have defined the integrals
\begin{equation}\label{Knq_int}
K_{n,q} \equiv \frac{1}{(2q\!+\!1)!!} \!\int\! \mathrm{dP}\,\tau_{\rm R}(x,p)  (u \cdot p)^{n-2q} (\Delta_{\alpha\beta} p^\alpha p^\beta)^q \, f_{\rm eq} \,. 
\end{equation}
We will now derive the second-order constitutive relation (and evolution equation) for the shear stress tensor in the next section.

%========================================================
\section{Second-order hydrodynamics} \label{sec:2ndordr}
\vspace{-.2cm}
%========================================================

The non-equilibrium correction to the distribution function till second order can be written as 
\begin{equation}
    f= f_{\rm eq} + \delta f_{(1)} + \delta f_{(2)} + \mathcal{O}(\partial^3) = f_{\rm eq} + \Delta f_{(2)} + \mathcal{O}(\partial^3),
\end{equation}
where we define $\Delta f_{(2)} \equiv \delta f_{(1)} + \delta f_{(2)}$ representing the non-equilibrium correction till second order. Using the kinetic equation~\eqref{ERTA}, and employing the Chapman-Enskog expansion, we obtain $\Delta f_{(2)}$ as, 
\begin{align}\label{deltaf2_exp}
    \Delta f_{(2)} = \Delta f^*_{(2)} 
    &- \frac{\taur}{u\cdot p} p^\mu \partial_\mu \delta f^*_{(1)} - \frac{\taur}{u\cdot p} p^\mu \partial_\mu f_{\rm eq}
\nonumber\\
    &+  \frac{\taur}{u\cdot p} p^\mu p^\nu  \partial_\mu \left( \frac{\taur}{u\cdot p} \partial_\nu f_{\rm eq} \right) .
\end{align}
Here $\delta f^*_{(1)}$ is out-of-equilibrium correction at first order and $\Delta f^*_{(2)}$ represents the correction up to second order.

As discussed in the previous section, the first-order contribution of $\delta u^\mu$ and $\delta T$ vanishes, and therefore they have contributions starting from second-order. Keeping terms till second-order in gradients, the first term on the right-hand side (r.h.s.) of the above equation is,
\begin{align}\label{deltaf2_1}
   \Delta f^*_{(2)} &= \left(-\frac{(\delta u \cdot p)}{T} + \frac{(u\cdot p)\delta T}{T^2} \right) f_{\rm eq}.
\end{align}
The second term on r.h.s. of Eq.~\eqref{deltaf2_exp},
\begin{align}\label{deltaf2_2}
   -\frac{\taur}{u\cdot p} p^\mu \partial_\mu \delta f^*_{(1)} = \mathcal{O}\left(\partial^3\right),
\end{align}
has correction starting from third-order in gradients because it involves derivatives of $\delta u^\mu$ and $\delta T$ which are at least second-order. The third term on the r.h.s. simplifies to,
\begin{align}\label{deltaf2_3}
    -\frac{\taur}{u\cdot p} p^\mu \partial_\mu &f_{\rm eq} = \frac{\taur}{T} \left[\frac{p^\mu p^\nu}{u\cdot p} \sigma_{\mu\nu} - 
        \frac{4}{3} \frac{u\cdot p}{\e+\p} \pi^{\mu\nu} \sigma_{\mu\nu} \right.
\nonumber\\
    &\ \left.- \frac{1}{\e+\p} \left( p^\mu \nabla_{\nu}\pi^{\nu}_{\mu} 
        - p^\mu  \pi^{\nu}_{\mu} \dot{u}_{\nu} \right) \right] f_{\rm eq}\,.
\end{align}
In deriving, we have kept all terms till second order when replacing derivatives of temperature with derivatives of fluid velocity using Eqs.~\eqref{beta_evol_1} and \eqref{beta_evol_2}. The last term on the r.h.s. of Eq.~\eqref{deltaf2_exp} is given by,

\begin{widetext}
    \begin{align} \label{deltaf2_4}
    \frac{\taur}{u\cdot p} p^\mu p^\nu  \partial_\mu \!\left(\! \frac{\taur}{u\cdot p} \partial_\nu f_{\rm eq} \!\right)\! =& 
    - \frac{\taur}{T} \left[ \dot{\tau}_{\rm R}\frac{p^\mu p^\nu}{u\cdot p} \sigma_{\mu\nu} 
    + (\nabla_\alpha \taur) \frac{p^\alpha p^\mu p^\nu}{(u\cdot p)^2} \sigma_{\mu\nu} \right] f_{\rm eq}
    - \frac{\taur^2}{T} \left[ \frac{2\theta}{3} \frac{p^\mu p^\nu}{u\cdot p} \sigma_{\mu\nu}
    +\frac{p^\mu p^\nu}{u\cdot p} \dot{\sigma}_{\mu\nu} \right.
\nonumber\\
    &\left. +\,\frac{p^\alpha p^\mu p^\nu }{(u\cdot p)^2} (\nabla_\alpha \sigma_{\mu\nu}) 
    -2 \frac{p^\alpha p^\mu p^\nu}{(u\cdot p)^2} \sigma_{\mu\nu} \dot{u}_\alpha 
    - \left(\frac{1}{T} \!+\! \frac{1}{u\cdot p} \right)\! \frac{(p^\mu p^\nu \sigma_{\mu\nu})^2}{(u\cdot p)^2} \right] f_{\rm eq} \,.
\end{align}
Therefore, the complete non-equilibrium correction till second order from Eqs.~\eqref{deltaf2_exp}-\eqref{deltaf2_4} is given by,
\begin{align}\label{delta_f2}
    \Delta f_{(2)} \!=\!& \left[\! \frac{(u\!\cdot\! p)\delta T}{T^2} \!-\! \frac{(\delta u \!\cdot\! p)}{T} 
    \!-\! \frac{\taur}{T} \!\left\{\! \left(\dot{\tau}_{\rm R}\!-\!1\right) \frac{p^\mu p^\nu}{u\!\cdot\! p} \sigma_{\mu\nu} 
    \!+\! \left(\nabla_\alpha \taur\right)\! \frac{p^\mu p^\nu p^\alpha}{(u\!\cdot\! p)^2} \sigma_{\mu\nu} \!+\! \frac{1}{\e\!+\!\p} \!\left(\! \frac{4}{3} (u\!\cdot\! p) \pi^{\mu\nu} \sigma_{\mu\nu} \!+ p^\mu \nabla_{\nu}\pi^{\nu}_{\mu} \!- p^\mu \pi^{\nu}_{\mu} \dot{u}_{\nu} \!\right) \!\!\right\} \right.
\nonumber\\
    &\ - \frac{\taur^2}{T} \!\left\{\! \frac{2\,p^\mu p^\nu}{3(u\!\cdot\! p)}\theta \sigma_{\mu\nu}\!
    +\frac{p^\mu p^\nu}{u\!\cdot\! p} \dot{\sigma}_{\mu\nu}\! 
    +\frac{p^\alpha p^\mu p^\nu }{(u\!\cdot\! p)^2} (\nabla_\alpha \sigma_{\mu\nu}\!) \right.
    - \left. \left. 2 \frac{p^\alpha p^\mu p^\nu}{(u\cdot p)^2} \sigma_{\mu\nu} \dot{u}_\alpha 
    \!-\! \!\left(\! \frac{1}{T} \!+\! \frac{1}{u\!\cdot\! p} \!\right)\! \frac{(p^\mu p^\nu \sigma_{\mu\nu})^2}{(u\cdot p)^2} \!\right\} \!\right]\! f_{\rm eq} \,.
\end{align}
\end{widetext}

%========================================================
\subsection{Imposing Landau frame conditions}\label{sec:lfc}
\vspace{-.2cm}
%========================================================

We note that the second-order correction to the equilibrium distribution function given by Eq.~\eqref{delta_f2} has the undetermined quantities $\delta u^\mu$ and $\delta T$. We determine these by imposing the Landau frame condition $\left(u_\mu T^{\mu\nu} = \e u^\nu\right)$ and matching condition $\left(\e = \e_0\right)$ with $f\mapsto f_{(2)}= f_{\rm eq} + \Delta f_{(2)}$ (see Appendix~\ref{app:LFC} for derivation),
\begin{align}\label{delta_u2}
\delta u^\mu &= \frac{5 K_{3,2}}{T(\e+\p)^2} \left(\pi^{\mu\nu} \dot{u}_\nu - \nabla_\nu \pi^{\mu\nu} -\pi^{\alpha\beta} \sigma_{\alpha\beta} u^\mu \right)
\nonumber\\
    &+ \frac{2L_{3,2}}{T(\e+\p)} \left(2\sigma^{\mu\nu} \dot{u}_\nu + \nabla_\nu \sigma^{\mu\nu} + \sigma^{\alpha\beta} \sigma_{\alpha\beta} u^\mu \right) ,
\end{align}
\begin{equation}\label{delta_T2}
\delta T = \frac{5}{3} \frac{K_{3,2}}{(\e+\p)^2} \pi^{\mu\nu} \sigma_{\mu\nu}
    + \frac{1}{\e+\p} \!\left(\! L_{3,2} -\frac{L_{4,2}}{3 T} \!\right)\! \sigma^{\mu\nu} \sigma_{\mu\nu} \,.
\end{equation}
The $L_{n,q}$ integrals appearing in the above expressions are defined as, 
\begin{equation}\label{Lnq_int}
    L_{n,q} \equiv \frac{1}{(2q\!+\!1)!!} \!\int\! \mathrm{dP}\,\tau_{\rm R}^2(x,p)  (u \cdot p)^{n-2q} (\Delta_{\alpha\beta} p^\alpha p^\beta)^q \, f_{\rm eq}\,.
\end{equation}
In the derivation, we have used the relation between the integrals,
\begin{equation}
    X_{n,q}= -\left(\frac{1}{2q+1}\right) X_{n,q-1} \,,
\end{equation}
which holds for all integrals defined in this article. We note that when the relaxation time does not depend on particle energy, the ERTA approximation of the collision term reduces to the Anderson-Witting RTA approximation, and consequently $\delta u^\mu$ and $\delta T$ vanishes (see Appendix~\ref{app:LFC}).

%========================================================
\subsection{Verification of microscopic conservation}\label{sec:micr_cons}
\vspace{-.2cm}
%========================================================

To verify microscopic energy-momentum conservation up to the second order, we show that the first momentum-moment of the collision kernel is at least third-order in gradients. To this end, we consider the first moment of the collision kernel in the Boltzmann equation~\eqref{ERTA} and substitute $f\mapsto f_{(2)} = f_{\rm eq} + \Delta f_{(2)}$, where $\Delta f_{(2)}$ is given by Eq.~\eqref{delta_f2},
\begin{equation}\label{del-T^mn_2}
\int \mathrm{dP}\, p^\nu \textfrak{C}[f] = -\int \mathrm{dP}\, \frac{(u \cdot p)}{\tau_{\rm R}} p^\nu \left(\Delta f_{(2)}-\Delta f^*_{(2)}\right)\! . 
\end{equation}
Using the expression of $\Delta f_{(2)}-\Delta f^*_{(2)}$ from Eq.~\eqref{deltaf2_exp},
\begin{equation}\label{rhs}
    \!\int\!\! \mathrm{dP}\, p^\nu \textfrak{C}[f] \!=\!\!\!\int\!\! \mathrm{dP}\, p^\mu p^\nu \partial_\mu f_{\rm eq}
    \!-\!\!\int\!\! \mathrm{dP}\, p^\nu p^\alpha p^\beta \partial_\alpha\! \!\left(\!\!\frac{\taur}{u\!\cdot\! p} \partial_\beta f_{\rm eq} \!\!\right)\!.
\end{equation}
The first term in the right-hand-side of the above equations is simplified as
\begin{equation}\label{rhs1}
    \int \mathrm{dP}\, p^\mu p^\nu \partial_\mu f_{\rm eq} = \dot{u}_{\mu}  \pi^{\mu\nu} -\nabla_\mu \pi^{\mu\nu} .
\end{equation}
Similarly, the second term is simplified as 
\begin{align}\label{rhs2}
    &\!\int\! \mathrm{dP}\, p^\nu p^\alpha p^\beta \partial_\alpha \!\!\left(\!\!\frac{\taur}{u\cdot p} \partial_\beta f_{\rm eq} \!\!\right)\! = \partial_\alpha  \!\!\int\!\! \mathrm{dP} \frac{\taur}{u\cdot p} p^\nu p^\alpha p^\beta \partial_\beta f_{\rm eq} 
\nonumber\\
    &= -\partial_\mu \!\left(\! 2 \frac{K_{3,2}}{T} \sigma^{\mu\nu} \!\right)\! +\mathcal{O}\!\left(\partial^3\right)\!
    = \dot{u}_{\mu}  \pi^{\mu\nu} \!-\nabla_\mu \pi^{\mu\nu} \!+ \mathcal{O}\!\left(\partial^3\right)\!.
\end{align}
In the last step, we have used the first-order constitutive relation~\eqref{shear_NS}. Using Eqs.~\eqref{rhs1} and \eqref{rhs2} in Eq.~\eqref{rhs}, we obtain
\begin{equation}
    \int \mathrm{dP}\, p^\nu \textfrak{C}[f] = \mathcal{O}\left(\partial^3\right).
\end{equation}
This demonstrates the preservation of microscopic energy-momentum conservation up to second order. It is noteworthy that $\delta u^\mu$ and $\delta T$ did not appear in the equations during the verification of microscopic conservation. This outcome is specific to the case of massless and chargeless particles and does not happen in general. The contribution from these quantities becomes essential to ensure the conservation of energy-momentum and net current in systems involving massive or charged particles, or at higher orders. We show that nontrivial cancellations due to these terms are necessary to ensure microscopic conservation till third order in Appendix~\ref{app:MCO3}.

%========================================================
\subsection{Shear stress till second-order}\label{sec:2ndorder}
\vspace{-.2cm}
%========================================================

The expression for shear stress tensor till second order in terms of the hydrodynamic fields is obtained by integrating $\Delta f_{(2)}$ in definition \eqref{SST},
\begin{align}\label{shear_2nd_ord}
    \pi^{\mu\nu} =&~ 2 \eta \sigma^{\mu\nu} - 2 \eta \tau_{\pi} \!\left( \dot{\sigma}^{\langle\mu\nu\rangle} + \frac{1}{3} \sigma^{\mu\nu} \theta \right) - \frac{4}{7} \frac{L_{4,2}}{T^2} \sigma_{\gamma}^{\langle\mu} \sigma^{\nu \rangle \gamma} 
\nonumber\\
    &+ \frac{4 L_{3,2}}{T} \sigma_{\gamma}^{\langle\mu} \omega^{\nu\rangle \gamma}\,,
\end{align}
where $\eta = K_{3,2}/T$ is the first order transport coefficient and we have defined $\tau_\pi \equiv L_{3,2}/K_{3,2}$. The equation presented above is consistent with the one derived in Ref.~\cite{Baier:2007ix} under the assumption of conformal symmetry. It is worth noting that the above equation retains its conformal invariance regardless of the specific functional dependence of the relaxation time on the particle energy.

One can rewrite Eq.~\eqref{shear_2nd_ord} as a relaxation-type equation for the evolution of shear stress tensor by replacing $\sigma^{\mu\nu} \to \pi^{\mu\nu}/(2 K_{3,2}/T)$\footnote{%
    In deriving, we used the relation
    \begin{equation*}
        \dot{\sigma}^{\langle\mu\nu\rangle} = \frac{\dot{\pi}^{\langle\mu\nu\rangle}}{2 \eta} -\left(\frac{T K_{3,2}+ Q_{3,2}- K_{4,2}}{3 \eta T^2}\right) \sigma^{\mu\nu}\theta \,,
    \end{equation*}
    where the $Q_{n,q}$ integral is defined as
    \begin{equation*}
        Q_{n,q} \equiv \frac{1}{(2q+1)!!} \int \mathrm{dP} \frac{\partial \tau_{\rm R}}{\partial \beta}(u \cdot p)^{n-2q} (\Delta_{\alpha\beta} p^\alpha p^\beta)^q f_{\rm eq} .
    \end{equation*}
    Further, we used the relation, $Q_{n,q} = K_{n+1,q} -(n+1) T K_{n,q} \,,$ and expressed $Q_{n,q}$ in terms of $K_{n,q}$ integral.
},
\begin{equation}\label{shear_evol2}
    \dot{\pi}^{\langle\mu\nu\rangle} +\frac{\pi^{\mu\nu}}{\tau_\pi} = 2 \beta_\pi \sigma^{\mu\nu} \!-\frac{4}{3} \pi^{\mu\nu} \theta + 2 \pi_\gamma^{\langle\mu}\omega^{\nu\rangle\gamma} \!- \mathcal{C} \pi_\gamma^{\langle\mu}\sigma^{\nu\rangle\gamma} \,,
\end{equation}
where $\beta_\pi \equiv \eta/\tau_\pi = \frac{\left(K_{3,2}\right)^2}{T L_{3,2}}$ and $\mathcal{C} \equiv \frac{2}{7}\frac{L_{4,2}}{T L_{3,2}}$. It is straightforward to verify that when the relaxation time is independent of particle energies, $\tau_\pi \to \taur, \, \beta_\pi \to (\e+\p)/5,$ and $\mathcal{C} \to 10/7$, which agrees with the previous results~\cite{Denicol:2010xn, Jaiswal:2013npa}. It is interesting to note that there is one new integral $K_{3,2}$ (corresponding to $\eta$) in first-order, and two new integrals, $L_{3,2}$ and $L_{4,2}$ (corresponding to $\tau_\pi$ and $\mathcal{C}$, respectively), in the second order.

As an illustration, we shall consider the following parametrization of the relaxation time~\cite{Dusling:2009df, Chakraborty:2010fr, Dusling:2011fd, Teaney:2013gca, Kurkela:2017xis}, 
\begin{equation}\label{tauR_parametrization}
    \taur(x,p) = \tau_{\rm eq}(x) \left(\frac{u\cdot p}{T}\right)^\ell, 
\end{equation}
where $\tau_{\rm eq}(x)$ represents the particle energy-independent part of relaxation time and scales as $1/T$ for conformal systems. We consider $\tau_{\rm eq}(x) = \kappa/T$, where $\kappa$ is a dimensionless constant. Note that the exponents $\ell$ may depend on the space-time coordinates. 
With this parametrization, the coefficient of shear viscosity is obtained to be \cite{Dash:2021ibx}, 
\begin{equation}\label{pl_eta}
    \eta = \frac{K_{3,2}}{T} = \frac{4\kappa T^3}{5\pi^2} \left[\frac{\Gamma(5+\ell)}{24} \right] ,
     \text{ iff } \ell >-5.
\end{equation}
Also, the coefficients $\tau_\pi$, $\beta_\pi$, and $\mathcal{C}$ appearing in Eq.~\eqref{shear_evol2} can be determined analytically to have the form,
\begin{align}\label{pl_2ndcoeff}
    \beta_\pi &\equiv \frac{\left(K_{3,2}\right)^2}{T L_{3,2}} = \frac{4 T^4}{5\pi^2} \left[\frac{\Gamma (\ell+5)^2}{24 \Gamma (2 \ell+5)} \right] ,
    \nonumber\\
    \tau_\pi &\equiv \frac{L_{3,2}}{K_{3,2}} = \frac{\kappa}{T} \left[ \frac{\Gamma (2 \ell+5)}{\Gamma (\ell+5)}\right] ,
    \nonumber\\
    \mathcal{C} &\equiv \frac{2}{7} \frac{L_{4,2}}{T L_{3,2}} = \frac{10+4\ell}{7},
\end{align}
with the condition $\ell >-5/2$. The above results will be employed in the next section to study the evolution of a plasma undergoing boost-invariant expansion.

%========================================================
\section{Bjorken flow}\label{sec:bjorken}
%========================================================

We shall now study the hydrodynamic equation obtained for a fluid undergoing Bjorken expansion~\cite{Bjorken:1982qr}. Bjorken symmetries enforce translational and rotational symmetry in the transverse $(x,y)$ plane, boost invariance along the $z$ (longitudinal) direction, and reflection symmetry $z \to -z$. The symmetries are manifest in Milne coordinate system ($\tau, x, y, \eta_s$), where $\tau= \sqrt{t^2 - z^2}$ is the proper time and $\eta_s=\tanh^{-1}\left(z/t\right)$ the space-time rapidity. In these coordinates the fluid appears to be static, $u^\mu = (1,0,0,0)$, irrotational ($\omega^{\mu\nu} = 0$) and unaccelerated ($\dot u^\mu = 0$), but has a non-zero local expansion rate, $\theta=1/\tau$. Symmetries further constrain the shear tensor to be diagonal and space-like in Milne coordinates, leaving only one independent component which we take to be the $\eta_s\eta_s$ component: $\pi^{xx}=\pi^{yy}= -\tau^2 \pi^{\eta_s\eta_s}/2 \equiv \pi/2$.

The hydrodynamic equations for evolution of energy density \eqref{evol1} and the shear tensor \eqref{shear_evol2} in Milne coordinates takes the form,
\begin{align}
    \frac{d\e}{d\tau} &=-\frac{1}{\tau}\left(\e+\p-\pi\right) ,
\label{eng_BJ}\\
    \frac{d\pi}{d\tau}&=-\frac{\pi}{\tau_\pi} +  \frac{4}{3} \frac{\beta_\pi}{\tau} - \left(\frac{4+\mathcal{C}}{3}\right) \frac{\pi}{\tau} \,.
\label{shear_BJ}
\end{align}
The above equations can be transformed into an equation for the quantity \cite{Blaizot:2017ucy, Blaizot:2019scw, Blaizot:2021cdv},
\begin{equation}
    g \equiv \frac{\tau}{\e} \frac{\partial \e}{\partial \tau} = \frac{\pi}{\e} - \frac{4}{3} \,.
\end{equation}
When the energy density exhibits power law behavior, $g$ corresponds to the exponent of that specific power law (i.e., if $\e \sim \tau^a$, then $g=a$). Equations~\eqref{eng_BJ} and \eqref{shear_BJ} can be written as a non-linear, first-order, differential equation in $g$ as,
\begin{equation}\label{eq_Bevol}
    -\mathcal{B}(g) \!=\! g^2 \!+ \!\left(\!\frac{8\!+\!\mathcal{C}}{3}\!\right)\! g 
    +\! \frac{4}{3} \!\left(\! \frac{4}{3} \!+\! \frac{\mathcal{C}}{3} \!-\! \frac{\beta_\pi}{\e} \!\!\right)\! 
    + \frac{\tau}{\tau_\pi}\!\left(\! g \!+\! \frac{4}{3} \!\right)\!,
\end{equation}
where we have defined $\mathcal{B}(g) \equiv \tau (\mathrm{d} g/\mathrm{d} \tau)$. Note that $\mathcal{C}$ and $\beta_\pi/\e$ are dimensionless.

%--------------------------------
\begin{figure}[t!]
    \centering
    \includegraphics[width=0.9\linewidth]{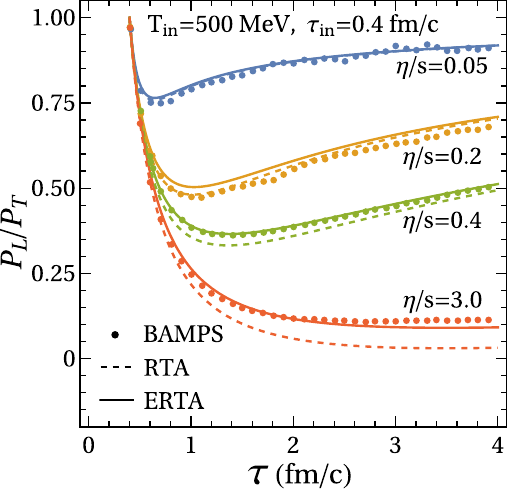}
    \vspace{-2mm}
    \caption{Evolution of $P_L/P_T$ in BAMPS (black dots) compared with hydrodynamic evolution using transport coefficients obtained from the RTA approximation (dashed curves) and ERTA approximation with $\ell =1/2$.}
    \vspace*{-2mm}
    \label{fig_1}
\end{figure}
%--------------------------------

%--------------------------------
\begin{figure*}[t!]
    \centering
    \includegraphics[width=\textwidth]{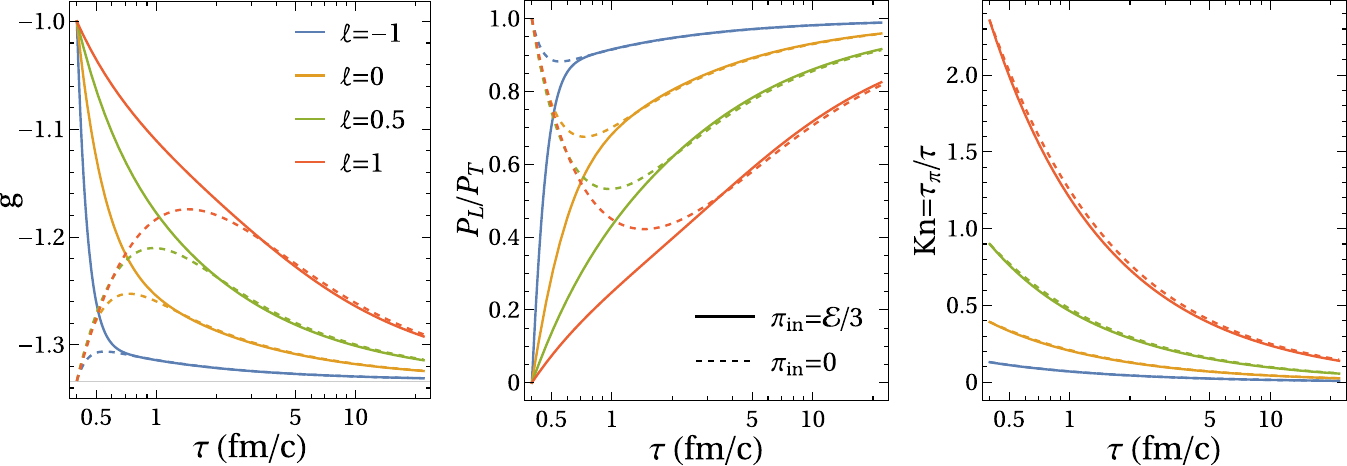}
    \vspace{-6mm}
    \caption{Time evolution of the quantity $g$, pressure anisotropy $P_L/P_T$, and Knudsen number $\tau_\pi/\tau$. Solid and dashed curves correspond to different initial shear stresses, $\pi = \e/3, 0$, respectively. The blue, orange, green, and red curves correspond to values of $\ell = -1, 0, 1/2, 1$, respectively. As can be seen in all three panels, isotropization and thermalization are delayed as the value of $\ell$ is increased.}
    \vspace*{-2mm}
    \label{fig_2}
\end{figure*}
%--------------------------------

The hydrodynamic regime is reached when the scattering rate exceeds the expansion rate, i.e., $\tau_\pi \ll \tau$. The last term in the above equation is dominant in this regime, and the hydrodynamic fixed point  $g_*$ is given by:
\begin{equation}
    g_* = - \frac{4}{3} \,.
\end{equation}
In the collisionless regime, the expansion rate far exceeds the scattering rate ($\tau_\pi \gg \tau$), and the function $\mathcal{B}(g)$ in Eq.~\eqref{eq_Bevol} is dominated by the terms that do not depend on $\tau/\tau_\pi$. The zeros of this function correspond to the free-streaming fixed points,
\begin{equation}\label{eq_FP}
    g_{\rm fp} = -\left(\frac{8+\mathcal{C}}{6} \right) \pm \sqrt{\frac{4\beta_\pi}{3\e} + \frac{\mathcal{C}^2}{36}} \,,
\end{equation}
with the positive root corresponding to the free-streaming stable fixed point. For a plasma undergoing Bjorken expansion, it has been shown that the stable free-streaming fixed point of the exact kinetic solution corresponds to vanishing longitudinal pressure, or $g=-1$ \cite{Chattopadhyay:2021ive, Jaiswal:2021uvv, Blaizot:2019scw}\footnote{%
    Although this was shown for the RTA Boltzmann equation, it holds true even for the ERTA case since the collision term vanishes in free-streaming.}. 
Using the parametrization~\eqref{tauR_parametrization} for the relaxation time and the corresponding values of the transport coefficients given in Eqs.~\eqref{pl_eta} and \eqref{pl_2ndcoeff}, we observe that the value of the stable fixed point in the exact kinetic equation ($g=-1$) can be recovered from Eq.~\eqref{eq_FP} for $\ell \approx 0.763$. Extending the domain of Israel-Stewart-type hydrodynamic theories requires the hydrodynamic equations to accurately capture the location of the stable free-streaming fixed point, as emphasized in Ref.~\cite{Jaiswal:2022udf}. Therefore, the evolution Eq.~\eqref{eq_Bevol}, or analogously Eqs.~\eqref{eng_BJ} and \eqref{shear_BJ}, is expected to provide a good description of the underlying weakly coupled microscopic theory with $\ell \approx 0.763$, even in far-off-equilibrium regimes. It is worth mentioning that this value of $\ell$ is not arbitrary; many microscopic theories lie in the range $\ell =[0,1]$ \cite{Dusling:2009df}.

To illustrate the impact of the ERTA framework, we show the comparison of the second-order hydrodynamic equations obtained from the RTA approximation ($\ell =0$) with those derived from the ERTA approximation setting $\ell =0.5$ and compare them with BAMPS results~\cite{Xu:2004mz, Xu:2007jv, El:2009vj} in Fig~\ref{fig_1}. The initial temperature is set to be $500\,$MeV at an initial time of $0.4\,$fm/c with a vanishing initial shear stress. Further, we fix the values of $\kappa$ appearing in Eqs.~\eqref{pl_eta} and \eqref{pl_2ndcoeff} such that $\eta/s$ is set to different values as mentioned in the figure\footnote{%
    For RTA approximation ($\ell=0$), $\kappa = \{1/4, 1, 2, 15\}$ for $\eta/s = \{0.05, 0.2, 0.4, 3.0\}$, respectively. Similarly for ERTA approximation with $\ell=0.5$, $\kappa = \{0.11, 0.46, 0.92, 6.9\}$ for $\eta/s = \{0.05, 0.2, 0.4, 3.0\}$, respectively.}. 
As can be seen from the figure, the solid curves representing the ERTA approximation with $\ell =0.5$ are in an overall better agreement with the BAMPS solution than the dashed curves~\cite{Mitra:2020gdk}\footnote{%
    We note that the second-order hydrodynamic equations obtained from the RTA approximation (dashed curve) perform better than the one obtained from ERTA approximation (solid curve) for $\eta/s=0.2$.
}.

In Figure~\ref{fig_2}, we present the evolution of three quantities: $g$, the pressure anisotropy $P_L/P_T = (P-\pi)/(P+\pi/2)$, and the Knudsen number $\tau_\pi/\tau$. The initial temperature is set to be $500\,$MeV at an initial time of $0.4\,$fm/c. Additionally, we consider the parameter $\kappa$ appearing in Eqs.~\eqref{pl_eta} and \eqref{pl_2ndcoeff} to have the value $5/(4\pi)$\footnote{%
    The value $\kappa= 5/(4\pi)$ implies $\eta/s=1/(4\pi)$ when the relaxation time is independent of the particle energies ($\ell=0$), i.e. when ERTA reduces to Anderson-Witting RTA.}. 
In all three panels, the solid curves represent cases where $P_L$ is initialized at $0$, corresponding to $\pi=\e/3$, while the dashed curves are initialized with a vanishing initial shear stress, $\pi =0$. The blue, orange, green, and red curves correspond to different values of $\ell = -1, 0, 1/2, 1$, respectively.
The left panel displays the evolution of the quantity $g$, with the gray solid line representing the hydrodynamic fixed point $g_*$. It can be observed from the systematic trend of blue, orange, green, and red curves that the system remains out of equilibrium for a longer duration as the values of $\ell$ are increased. This feature is also visible in the middle panel where the evolution of pressure anisotropy, $P_L/P_T$ is shown -- approach to $P_L/P_T =1$ is delayed for the orange, green, and red curves compared to the blue curve, indicating a slower isotropization. Also, in the left and middle panels we observe that the solid and dashed curves, representing different initial shear stress, overlap earlier for smaller values of $\ell$. Interestingly, the evolution of Knudsen number shown in the right panel is not strongly dependent on the initial values of shear stress but has a strong dependence on the strength of the momentum-dependence of the relaxation time i.e. on $\ell$; the solid and dashed curves largely overlap during the entire evolution. It is worth noting that increasing the value of $\ell$ enhances the initial gradient strength (as $\tau_\pi$ increases), and smaller values of $\ell$ drive the system towards thermalization at a faster rate, which is evident from the middle panel.

\begin{figure}[t!]
    \centering
    \includegraphics[width=\linewidth]{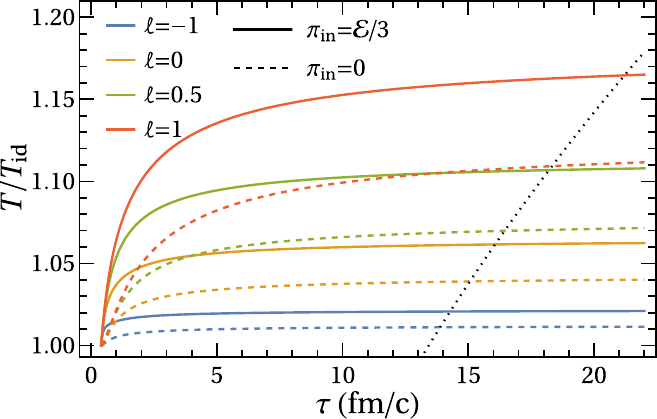}
    \vspace{-6mm}
    \caption{Time evolution of temperature normalized with ideal temperature evolution. The initial conditions and legends are the same as in Fig.~\ref{fig_2}. The black dotted curve represents a temperature surface of $155\,$MeV.}
    \vspace*{-2mm}
    \label{fig_3}
\end{figure}

\begin{figure}[t!]
    \centering
    \includegraphics[width=\linewidth]{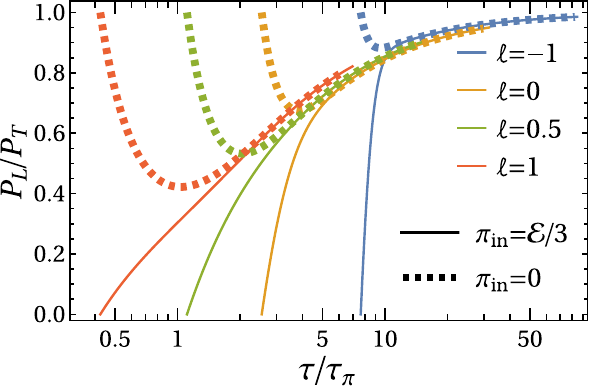}
    \vspace{-6mm}
    \caption{Evolution of pressure anisotropy $P_L/P_T$ with $\tau/\tau_\pi$. The evolution of different curves is stopped at proper times when the temperature evolution reaches $155\,$MeV.}
    \vspace*{-2mm}
    \label{fig_4}
\end{figure}

In Figure~\ref{fig_3}, we show the evolution of the temperature normalized with the ideal temperature evolution, $T_{\rm id} = T_{\rm in} (\tau_{\rm in}/\tau)^{1/3}$. It is observed that at a given time, the fluid maintains a higher temperature when the initial shear stress has a large positive value (solid curves), in contrast to when the initial shear stress is vanishing (dashed curves). An interesting observation is that increasing values of $\ell$ also lead to higher temperatures of the medium, as indicated by the trend of the differently colored curves. This may be understood from the right panel of Fig.~\ref{fig_2}, where we observe that an increase in the values of $\ell$ results in a larger Knudsen number. Consequently, this leads to increased dissipation, resulting in a slower fall of temperature compared to ideal evolution. Moreover, the interplay between the initial conditions for shear stress and the various medium interactions (characterized by different values of $\ell$) is intriguing, and can provide insights towards constraining the initial conditions for hydrodynamic simulation of heavy-ion collisions. Further, the various curves in Fig.~\ref{fig_3} crossing the temperature surface of $155\,$MeV (represented by the black dotted curve) at different proper times suggests that a constant temperature particlization surface can be reached at different times with varying anisotropies. This can be seen  more clearly in Fig.~\ref{fig_4}, where the evolution of the pressure anisotropy $P_L/P_T$ with $\tau/\tau_\pi$ is shown. The evolution of the curves is stopped when the temperature of the plasma reaches $155\,$MeV during the expansion (at times when the different curves cross the black dotted curve in Fig.~\ref{fig_3}). In Fig.~\ref{fig_4}, we see that the pressure anisotropy across the different colored curves differs significantly. We also note that the evolution of $P_L/P_T$ for different curves in the near-equilibrium regime ($\tau\gtrsim 5\tau_\pi$) is nearly the same, but differs substantially in the far-off-equilibrium regime.

%--------------------------------
%--------------------------------

%========================================================
\section{Summary and outlook}\label{sec:summary}
%========================================================

To summarize, we have derived relativistic second-order hydrodynamics from the Boltzmann equation using the extended relaxation time approximation for the collision kernel, incorporating an energy-dependent relaxation time. The transport coefficients are shown to explicitly depend on the microscopic relaxation rate. We investigated the fixed point structure of the hydrodynamic equations for a plasma undergoing Bjorken flow and showed that the location of the free-streaming fixed points depends on the energy dependence of the relaxation time. Additionally, we employed a power law parametrization to describe the energy dependence of the relaxation time and examined its impact on the thermalization process of the expanding plasma. We demonstrated that the plasma's approach to equilibrium is affected by the relaxation time's dependence on different powers of energy; the plasma remains in the out-off-equilibrium regime and at a higher temperature for longer duration as larger positive values of $\ell$ are considered.

%--------------------------------
%--------------------------------

While the derivation in the present article is done for a conformal system without conserved charges, it can be extended for non-conformal systems with conserved charges and quantum statistics by following the steps outlined in the article. It is also desired to have typical relaxation rates for the energy dependence of the relaxation time across different stages of the evolution of the nuclear matter formed in heavy-ion collisions. Such parametrization of the relaxation time can have parameters which may depend, for example, on the temperature of the medium\footnote{%
    It remains to be explored if some of the essential features of a strongly coupled fluid can be captured in this framework by parametrizing the relaxation time.
}. Incorporating these rates will make the full hydrodynamic equations with the associated transport coefficients more suitable for a (3+1)-dimensional hydrodynamic simulation. It should be noted that the functional form of the first-order transport coefficients, such as $\eta$, is determined within the framework. Furthermore, such an analysis may also provide insights into the form of distribution function at particlization. These aspects will be investigated in future studies.

%%%%%%%%%%%%%%%%%%%%%%%%%%%%%%%%%%%%%%
\acknowledgments

S.J. thanks Richard J. Furnstahl and Ulrich Heinz for insightful comments and suggestions on the manuscript. He acknowledges the kind hospitality of NISER, Bhubaneswar where part of this work was done. A.J. is supported in part by the DST-INSPIRE faculty award under Grant No. DST/INSPIRE/04/2017/000038. S.J. is supported by the NSF CSSI program under award no. OAC-2004601 (BAND Collaboration). S.B. kindly acknowledges the support of the Faculty of Physics, Astronomy and Applied Computer Science, Jagiellonian University Grant No. LM/17/SB.

%%%%%%%%%%%%%%%%%%%%%%%%%%%%%%%%%%%%%%

%==========================================
\appendix
%==========================================

%%%%%%%%%%%%%%%%%%%%%%%%%%%%%%%%%%%%%%%%%%%%%
\section{Derivation of \texorpdfstring{$\delta u^\mu$ and $\delta T$}{}} \label{app:LFC}
%%%%%%%%%%%%%%%%%%%%%%%%%%%%%%%%%%%%%%%%%%%%%

In this Appendix, we obtain $\delta u^\mu$ and $\delta T$ by imposing Landau frame and matching conditions. The Landau Frame condition, $u_{\mu} T^{\mu \nu} =\e u^{\nu}$, with the matching $\e = \e_0$, for a non-equilibrium distribution $f=f_{\rm eq} + \delta f$ can be written as 
\begin{align}
     & u_\mu \int \mathrm{dP}\, p^\mu p^\nu (f_{\rm eq} +\delta f) = u^\nu \int \mathrm{dP}\, (u\cdot p)^2 f_{\rm eq}
\nonumber \\
    \implies &  u_\mu \int \mathrm{dP}\, p^\mu p^\nu \delta f = u^\nu u_\alpha u_\beta I^{\alpha\beta} - u_\mu I^{\mu\nu} = 0,
\end{align}
where $I^{\alpha\beta} = \int \mathrm{dP}\, p^\alpha p^\beta f_{\rm eq}$. Replacing $\delta f \mapsto \delta f_{(2)}$ obtained in Eq.~\eqref{delta_f2}, and performing the integrals in the local rest frame of $u^\mu$, the above equation reduces to
\begin{align}\label{LFC_res_1}
&I_{3,1} \delta u^\mu - I_{3,0} \frac{\delta T}{T} u^\mu + \frac{5 K_{3,2}}{(\e+\p)} \left(\pi^{\mu\nu}\dot{u}_\nu - \nabla_\nu \pi^{\mu\nu} \right)
\nonumber\\
    &+ \left(6 L_{3,2} + 2 M_{4,2} - 2 \frac{N_{3,2}}{T} \right) \sigma^{\mu\nu} \dot{u}_\nu + 2L_{3,2} \nabla_\nu \sigma^{\mu\nu}
\nonumber\\
    &+ \left( 10 L_{3,2} - 2\frac{L_{4,2}}{T} + 2M_{4,2} \right) \sigma^{\alpha\beta} \sigma_{\alpha\beta} u^\mu = 0 .
\end{align}
Note that the term $\delta u \cdot u = (\delta u\cdot \delta u)/2 \sim \mathcal{O}\left( \delta^4\right)$,  since $\delta u^\mu$ is at least second order (see discussion in Section~\ref{sec:2ndordr}), and has been ignored in the derivation. Further, we have defined the thermodynamic integrals,
\begin{align}
I_{n,q} \!&\equiv\! \frac{1}{(2q+1)!!} \!\int\! \mathrm{dP} (u \cdot p)^{n-2q} (\Delta_{\alpha\beta} p^\alpha p^\beta)^q \, f_{\rm eq} ,
\\
M_{n,q} \!&\equiv\! \frac{1}{(2q\!+\!1)!!} \!\int\!\! \mathrm{dP} \tau_{\rm R} \frac{\partial \tau_{\rm R}}{\partial (u\!\cdot\! p)} (u \!\cdot\! p)^{n-2q} (\Delta_{\alpha\beta} p^\alpha p^\beta)^q f_{\rm eq} ,
\\
N_{n,q} \!&\equiv\! \frac{1}{(2q\!+\!1)!!} \!\int\! \mathrm{dP}\, \tau_{\rm R} \frac{\partial \tau_{\rm R}}{\partial \beta} (u\! \cdot\! p)^{n-2q} (\Delta_{\alpha\beta} p^\alpha p^\beta)^q \, f_{\rm eq} .
\end{align}
We note that $M_{n,q}$ and $N_{n,q}$ integrals can be expressed in terms of the $L_{n,q}$ integrals through the relations,
\begin{align}
M_{n,q} &= \frac{1}{2T} L_{n,q} - \frac{n+1}{2} L_{n-1,q} \,, \quad \text{iff } n > -1\,,
\\
2 N_{n,q} &= L_{n+1,q} - n T L_{n,q} \,.
\end{align}
Using these relations, Eq.~\eqref{LFC_res_1} simplifies to,
\begin{align}\label{LFC_res}
&I_{3,1} \delta u^\mu \!- I_{3,0} \frac{\delta T}{T} u^\mu \!+ \frac{5 K_{3,2}}{(\e+\p)} \left( \pi^{\mu\nu}\dot{u}_\nu \!- \nabla_\nu \pi^{\mu\nu} \right) \!+ 2 L_{3,2}\nonumber\\
& \times\!\left( 2 \sigma^{\mu\nu} \dot{u}_\nu \!+\! \nabla_\nu \sigma^{\mu\nu} \right) + \!\left(\! 5 L_{3,2} \!-\! \frac{L_{4,2}}{T} \!\right)\! \sigma^{\alpha\beta} \sigma_{\alpha\beta} u^\mu = 0 .
\end{align}

Similarly, using the matching condition: 
\begin{align}
    u_{\mu}u_{\nu} T^{\mu \nu} = \e_0
    \implies u_\mu u_\nu \int \mathrm{dP}\, p^\mu p^\nu \delta f = 0 \,,
\end{align}
and replacing $\delta f \mapsto \delta f_{(2)}$ obtained in Eq.~\eqref{delta_f2}, we obtain
\begin{equation}\label{LMC_res}
- I_{3,0} \frac{\delta T}{T} + \frac{5 K_{3,2}}{\e\!+\!\p} \pi^{\mu\nu} \sigma_{\mu\nu} + \!\left(\! 3 L_{3,2} -\frac{L_{4,2}}{T} \!\right)\! \sigma^{\mu\nu} \sigma_{\mu\nu}  = 0\,.
\end{equation}
Noting that $I_{3,0} = 3T (\e\!+\!\p)$ and solving for $\delta T$, we obtain
\begin{equation}\label{app_deltaT}
\delta T= \frac{5}{3} \frac{K_{3,2}}{(\e\!+\!\p)^2} \pi^{\mu\nu} \sigma_{\mu\nu}
 + \frac{1}{\e\!+\!\p} \!\left(\! L_{3,2} -\frac{L_{4,2}}{3 T} \!\right)\! \sigma^{\mu\nu} \sigma_{\mu\nu} \,.
\end{equation}
The expression for $\delta u^\nu$ is obtained by using Eq.~\eqref{LMC_res} in Eq.~\eqref{LFC_res},
\begin{align}\label{app_deltau}
\delta u^\mu &= \frac{5 K_{3,2}}{T(\e\!+\!\p)^2} \left(\pi^{\mu\nu} \dot{u}_\nu - \nabla_\nu \pi^{\mu\nu} -\pi^{\alpha\beta} \sigma_{\alpha\beta} u^\mu \right) \nonumber\\
&+ \frac{2L_{3,2}}{T(\e\!+\!\p)} \left(2\sigma^{\mu\nu} \dot{u}_\nu + \nabla_\nu \sigma^{\mu\nu} + \sigma^{\alpha\beta} \sigma_{\alpha\beta} u^\mu \right) , 
\end{align}
where we have used the relation $I_{3,1} = -T (\e\!+\!\p)$.

In the case when the relaxation time is independent of the particle energies, the integrals $K_{n,q} \to \taur\, I_{n,q}$ and $L_{n,q} \to \taur^2\, I_{n,q} = \taur K_{n,q}$. Using these, and noting that $\taur = 5 \eta/(\e\!+\!\p)$, Eq.~\eqref{app_deltau} simplifies to
\begin{align}\label{app_deltau_mi}
\delta u^\mu &= \frac{5 K_{3,2}}{T(\e\!+\!\p)^2} \left(\pi^{\mu\nu} \dot{u}_\nu - \nabla_\nu \pi^{\mu\nu} -\pi^{\alpha\beta} \sigma_{\alpha\beta} u^\mu \right) \nonumber\\
&+ \frac{5 K_{3,2}}{T(\e\!+\!\p)^2} 2\eta  \left(2\sigma^{\mu\nu} \dot{u}_\nu + \nabla_\nu \sigma^{\mu\nu} + \sigma^{\alpha\beta} \sigma_{\alpha\beta} u^\mu \right) \nonumber\\
&= \frac{5 K_{3,2}}{T(\e\!+\!\p)^2} \left(3 \pi^{\mu\nu} \dot{u}_\nu - \nabla_\nu \pi^{\mu\nu} + 2\eta\nabla_\nu \sigma^{\mu\nu} \right)\nonumber\\
&=0 \,.
\end{align}
In transitioning to the second equality, we employed the first-order relation $\pi^{\mu\nu} = 2 \eta \sigma^{\mu\nu}$. Furthermore, we used the relation $\nabla_\nu \eta = -3T\eta \nabla_\nu \beta = 3 \eta \dot{u}_\nu$ in the last equality.

Similarly, Eq.~\eqref{app_deltaT} reduces to
\begin{align}\label{app_deltaT_mi}
\delta T &= \frac{1}{\e\!+\!\p} \!\left[\frac{5}{3} \frac{K_{3,2}}{(\e\!+\!\p)} \pi^{\mu\nu}\! \sigma_{\mu\nu} \!+ \taur^2\! \left(\!\! I_{3,2} \!-\!\frac{5}{3} I_{3,2} \!\!\right)\! \sigma^{\mu\nu}\! \sigma_{\mu\nu}  \!\right] \nonumber\\
&= \frac{1}{\e\!+\!\p} \!\left[\frac{5}{3} \frac{K_{3,2}}{(\e\!+\!\p)} \pi^{\mu\nu} \sigma_{\mu\nu} - \frac{5\eta}{(\e\!+\!\p)} \frac{2}{3} K_{3,2} \sigma^{\mu\nu} \sigma_{\mu\nu} \right] \nonumber\\
&=0 \,.
\end{align}
The fact that $\delta u^\mu$ and $\delta T$ vanish when the relaxation time is independent of particle energy is anticipated since ERTA reduces to the Anderson-Witting RTA, thus providing a consistency validation.

\pagebreak
%%%%%%%%%%%%%%%%%%%%%%%%%%%%%%%%%%%%%%%%%%%%%%%%%
\section{Microscopic conservation at third order} \label{app:MCO3}
%%%%%%%%%%%%%%%%%%%%%%%%%%%%%%%%%%%%%%%%%%%%%%%%%

In this appendix, we demonstrate microscopic energy-momentum conservation up to third order. The non-equilibrium correction to the equilibrium distribution function up to third order in spacetime gradients from the Boltzmann equation is obtained to be,
\begin{align}
    \Delta f_{(3)} =& \Delta f^*_{(3)} - \frac{\taur}{(u\cdot p)} p^\mu \partial_\mu \left[ f_{\rm eq} - \frac{\taur}{(u\cdot p)} p^\rho \partial_\rho f_{\rm eq} \right.
\nonumber\\
    &+ \frac{\taur}{(u\cdot p)} p^\rho p^\nu \partial_\rho \left( \frac{\taur}{(u\cdot p)} \partial_\nu f_{\rm eq} \right) 
\nonumber\\
    &\left.  - \frac{\taur}{(u\cdot p)} p^\rho \partial_\rho \Delta f^*_{(1)} + \Delta f^*_{(2)} \right] .
    \label{Delta_f_3}
\end{align}
As discussed in Sec.~\ref{sec:micr_cons}, verifying microscopic energy-momentum conservation up to the third order amounts to showing that the first momentum-moment of the collision kernel vanishes. Substituting $f \mapsto f_{(3)} = f_{\rm eq} + \Delta f_{(3)}$, and using Eq.~\eqref{Delta_f_3},
\begin{align}
    &\int \mathrm{dP}\, p^\nu \mathfrak{C}[f] = -\int \mathrm{dP}\, \frac{(u \cdot p)}{\tau_{\rm R}} p^\nu \left(\Delta f_{(3)}-\Delta f^*_{(3)}\right)
\nonumber\\
    &= \underbrace{\int\! \mathrm{dP}\, p^\nu p^\mu \left(\partial_\mu f_{\rm eq}\right)}_{I}
    + \underbrace{\int\! \mathrm{dP}\, p^\nu p^\mu p^\rho \partial_\mu\left[-\frac{\taur}{(u\cdot p)} \left(\partial_\rho f_{\rm eq}\right)\right]}_{II}
\nonumber\\
    &+ \underbrace{\int\! \mathrm{dP}\, p^\nu p^\mu p^\rho p^\lambda \partial_\mu \left[\frac{\taur}{(u\cdot p)} \partial_\rho \left\{\frac{\taur}{(u\cdot p)} \left(\partial_\lambda f_{\rm eq}\right) \right\} \right]}_{III}
\nonumber\\
    &+ \underbrace{\int\! \mathrm{dP}\, p^\nu p^\mu p^\rho \partial_\mu \!\left[\frac{\taur}{(u\cdot p)} \!\left(\! \partial_\rho \Delta f^*_{(1)} \!\right)\! \right]}_{IV} 
    \!+\! \underbrace{\int\! \mathrm{dP}\, p^\nu p^\mu \!\left(\! \partial_\mu \Delta f^*_{(2)} \!\right)\! }_{V} .
    \label{consrv_o3}
\end{align}

Evaluating these integrals, we find,
\begin{align}
    \textit{I} =& \dot{u}_{\mu}  \pi^{\mu\nu} -\nabla_\mu \pi^{\mu\nu} = - \partial_\mu  \pi^{\mu\nu} \,, 
    \label{Int:1}
\\
    \textit{II} =& \partial_\mu \!\bigg[\! \frac{2 K_{3,2}}{T} \sigma^{\mu\nu}
    + \frac{5 K_{3,2}}{T(\e+\p)} \!\bigg\{\! \!\left(\! u^\mu u^\nu \!+\! \frac{\Delta^{\mu\nu}}{3} \!\right)\! \pi^{\alpha \beta} \sigma_{\alpha\beta} 
\nonumber\\
    &\quad \!+\! u^\mu \nabla_\gamma \pi^{\nu\gamma} + u^\nu \nabla_\gamma \pi^{\mu\gamma} 
    \!-\! \dot{u}_\gamma \left( u^\mu \pi^{\nu\gamma} + u^\nu \pi^{\mu\gamma} \right) \!\Big\}\! \bigg]. \label{Int:2}
\end{align}
In deriving, we have used the hydrodynamic evolution equations~\eqref{beta_evol_1} and \eqref{beta_evol_2}. The integral \textit{III}, consistently keeping all terms till third order in gradients, is obtained to be,
\begin{align}
    \textit{III} =& \partial_\mu \!\left[- \frac{2L_{3,2}}{T} \Big\{ \dot{\sigma}^{\mu\nu} \!+\! \frac{1}{3} \sigma^{\mu\nu} \theta + u^\mu \nabla_\rho \sigma^{\nu\rho} + u^\nu \nabla_\rho \sigma^{\mu\rho} \right.
\nonumber\\
    &\qquad + 3 \dot{u}_\rho \left( u^\mu \sigma^{\nu\rho} + u^\nu \sigma^{\mu\rho} \right) +  \omega^{\nu}_\rho \sigma^{\mu\rho} +\omega^{\mu}_\rho \sigma^{\nu\rho} \!\Big\} 
\nonumber\\
    &\qquad  + \left(\frac{L_{3,2}}{T} - \frac{1}{7} \frac{L_{4,2}}{T^2} \right)  \left( \Delta^{\mu\nu}-7 u^\mu u^\nu \right) \sigma_{\alpha\beta} \sigma^{\alpha\beta} 
\nonumber\\
    &\qquad \left. -\, \frac{4}{7} \frac{L_{4,2}}{T^2} \sigma^{\mu}_\rho \sigma^{\nu\rho} \right]  +\order{\partial^4} \,.
    \label{Int:3_1}
\end{align}
The above equation can be further simplified by using the second-order constitutive relation~\eqref{shear_2nd_ord} for the shear stress tensor which can be written as,
{\small 
\begin{align}
    \pi^{\mu\nu} &= \frac{2K_{3,2}}{T} \sigma^{\mu\nu} + \frac{4}{7} \frac{L_{4,2}}{T^2 } \left( \frac{1}{3} \Delta^{\mu\nu} \sigma_{\alpha\beta} \sigma^{\alpha\beta} - \sigma^{\mu}_\rho \sigma^{\nu\rho} \right)
    \nonumber\\
    & \!-\! \frac{2L_{3,2}}{T} \!\left(\! \dot{\sigma}^{\mu\nu} \!+\! \frac{1}{3}\sigma^{\mu\nu}\theta \!+\! \omega^{\nu}_\rho \sigma^{\mu\rho} \!+\! \omega^{\mu}_\rho \sigma^{\nu\rho} \!+\! \dot{u}_\rho ( u^\mu \sigma^{\nu\rho} \!+\! u^\nu \sigma^{\mu\rho} ) \!\right)\!.
    \label{app:shear_2nd_ord}
\end{align}
}
Replacing the $\dot{\sigma}^{\mu\nu}$ term in Eq.~\eqref{Int:3_1} using the above equation, integral \textit{III} simplifies to,
\begin{align}
    \textit{III} &=  \partial_\mu \left[ \pi^{\mu\nu} - \frac{2K_{3,2}}{T} \sigma^{\mu\nu} 
    + \frac{L_{4,2}}{T^2} \left( u^\mu u^\nu -\frac{\Delta^{\mu\nu}}{3} \right) \sigma^{\alpha \beta} \sigma_{\alpha\beta} \right.
    \nonumber\\
    &\qquad\quad  \!-\! \frac{2L_{3,2}}{T} \!\bigg\{\! u^\mu \nabla_\rho \sigma^{\nu\rho} \!+\! u^\nu \nabla_\rho \sigma^{\mu\rho} \!+\! 2\dot{u}_\rho ( u^\mu \sigma^{\nu\rho} + u^\nu \sigma^{\mu\rho})
    \nonumber\\
    &\qquad\qquad\qquad\quad \left. +\frac{1}{2} \left( 7 u^\mu u^\nu -\Delta^{\mu\nu}\right) \sigma^{\alpha \beta} \sigma_{\alpha\beta} \bigg\} \right] +\order{\partial^4}\,.
    \label{Int:3}
\end{align}
Note that the replacement using Eq.~\eqref{shear_2nd_ord} (or Eq.~\eqref{app:shear_2nd_ord}) keeps the integral \textit{III} exact up to third order. 

\begin{widetext}
Adding the contributions of the integrals \textit{I}, \textit{II} and \textit{III} using Eqs.~(\ref{Int:1}, \ref{Int:2}, \ref{Int:3}):
\begin{align}
    \textit{I} + \textit{II} + \textit{III} = \partial_\mu \bigg[& \frac{5 K_{3,2}}{T(\e+\p)} \left\{ \!\left(\! u^\mu u^\nu + \frac{\Delta^{\mu\nu}}{3} \!\right)\! \pi^{\alpha \beta} \sigma_{\alpha\beta} + u^\mu \nabla_\rho \pi^{\nu\rho}  + u^\nu \nabla_\rho \pi^{\mu\rho} 
    - \dot{u}_\rho \left( u^\mu \pi^{\nu\rho} + u^\nu \pi^{\mu\rho} \right) \right\}
\nonumber\\
    &- \frac{2L_{3,2}}{T} \left\{ u^\mu \nabla_\rho \sigma^{\nu\rho} + u^\nu \nabla_\rho \sigma^{\mu\rho} + 2\dot{u}_\rho ( u^\mu \sigma^{\nu\rho} + u^\nu \sigma^{\mu\rho})  +\frac{1}{2} \left( 7 u^\mu u^\nu -\Delta^{\mu\nu}\right) \sigma^{\alpha \beta} \sigma_{\alpha\beta} \right\}
\nonumber\\
    &+ \frac{L_{4,2}}{T^2} \left( u^\mu u^\nu -\frac{\Delta^{\mu\nu}}{3} \right) \sigma^{\alpha \beta} \sigma_{\alpha\beta} \bigg] +\order{\partial^4}\,.
    \label{Int_unstared}
\end{align}
It can be verified that the above combination vanishes when the relaxation time is independent of particle momenta. This is consistent with third-order hydrodynamics derived using the RTA approximation of the collision kernel~\cite{Jaiswal:2013vta}.
\end{widetext}

We now evaluate the two integrals, \textit{IV} and \textit{V}, which are due to the difference between the `thermodynamic' and hydrodynamic frames. Since $\Delta f^*_{(1)}$ contains the terms $\delta u^\mu,\, \delta T$ which are at least $\order{\partial^2}$, the integral \textit{IV},
\begin{equation}\label{Int:4}
    \textit{IV} = \!\int\! \mathrm{dP} p^\nu p^\mu p^\rho \partial_\mu \left[\frac{\taur}{(u\cdot p)} \left(\partial_\rho \Delta f^*_{(1)}\right)\right] = \order{\partial^4}
\end{equation}
is at least $\order{\partial^4}$ and can be ignored. The remaining integral \textit{V} simplifies to
\begin{align}
    \textit{V} &= \int \mathrm{dP}\, p^\nu p^\mu \left(\partial_\mu \Delta f^*_{(2)}\right) 
\nonumber\\
    &= \partial_\mu \!\left[\!-\frac{ I_{3,1}}{T} \left( u^\nu \delta u^\mu + u^\mu \delta u^\nu \right) \!+\! \frac{\delta T}{T^2} \left( u^\mu u^\nu I_{3,0} + \Delta^{\mu\nu} I_{3,1} \right) \!\right]\!
\nonumber\\
    &= \partial_\mu \!\left[\! (\e+\p) \!\left\{\! \left( u^\nu \delta u^\mu + u^\mu \delta u^\nu \right) \!+\! \frac{\delta T}{T} \left( 3 u^\mu u^\nu \!-\! \Delta^{\mu\nu} \right) \!\right\} \!\right]\!
    , \label{Int:5_1}
\end{align}
where we have not considered the terms $\left(u \cdot\delta u\right) = -(\delta u)^2/2,\, (\delta u)^2,\, (\delta T)^2,\, \delta u^\mu \delta T$, and higher order terms in the Taylor expansion of $\Delta f^*_{(2)}$ as they are at least $\order{\partial^4}$. Hence, the expansion of $\Delta f^*_{(2)}$ only involves contributions from the terms $\delta u^\mu$ and $\delta T$, and their determination imposing Landau frame and matching conditions remains identical as done for second-order in Appendix~\ref{app:LFC}. The relevant expressions for $\delta u^\mu$ and $\delta T$ can be found in Eqs.~\eqref{delta_u2} and \eqref{delta_T2}. It is important to note that when the relaxation time is independent of particle energies, both $\delta u^\mu$ and $\delta T$ vanish, as demonstrated in Appendix~\ref{app:LFC}, resulting in the integral~\textit{V} also vanishing.

Using these, the integral \eqref{Int:5_1} can be written as,
\begin{widetext}
\begin{align}   
    \textit{V} 
    =& \partial_\mu \left[ \frac{u^\nu}{T} \left\{ \frac{5 K_{3,2}}{(\e+\p)} \left(\pi^{\mu\rho} \dot{u}_\rho - \nabla_\rho \pi^{\mu\rho} - \pi^{\alpha\beta} \sigma_{\alpha\beta} u^\mu \right) + 2 L_{3,2} \left( 2 \sigma^{\mu\rho} \dot{u}_\rho + \nabla_\rho \sigma^{\mu\rho} + \sigma^{\alpha\beta} \sigma_{\alpha\beta} u^\mu \right) \right\} \right.
\nonumber\\
    &\qquad\quad + \frac{u^\mu}{T} \left\{ \frac{5 K_{3,2}}{(\e\!+\!\p)} \left(\pi^{\nu\rho} \dot{u}_\rho - \nabla_\rho \pi^{\nu\rho} - \pi^{\alpha\beta} \sigma_{\alpha\beta} u^\nu \right) + 2 L_{3,2} \left(2 \sigma^{\nu\rho} \dot{u}_\rho + \nabla_\rho \sigma^{\nu\rho} + \sigma^{\alpha\beta} \sigma_{\alpha\beta} u^\nu \right) \right\} 
\nonumber\\
    &\qquad\quad \left.+ \frac{1}{T} \left\{ \frac{5}{3} \frac{K_{3,2}}{(\e + \p)} \pi^{\alpha\beta} \sigma_{\alpha\beta} + \left( L_{3,2} -\frac{L_{4,2}}{3T} \right) \sigma^{\alpha\beta} \sigma_{\alpha\beta} \right\} \left( 3 u^\mu u^\nu \!-\! \Delta^{\mu\nu} \right) \right] +\order{\partial^4} \,,
\nonumber\\
 =&  -\partial_\mu \bigg[ \frac{5 K_{3,2}}{T(\e+\p)} \left\{ \!\left(\! u^\mu u^\nu + \frac{\Delta^{\mu\nu}}{3} \!\right)\! \pi^{\alpha \beta} \sigma_{\alpha\beta} + u^\mu \nabla_\rho \pi^{\nu\rho}  + u^\nu \nabla_\rho \pi^{\mu\rho} 
    - \dot{u}_\rho \left( u^\mu \pi^{\nu\rho} + u^\nu \pi^{\mu\rho} \right) \right\}
\nonumber\\
    &\qquad\quad - \frac{2L_{3,2}}{T} \left\{ u^\mu \nabla_\rho \sigma^{\nu\rho} + u^\nu \nabla_\rho \sigma^{\mu\rho} + 2\dot{u}_\rho ( u^\mu \sigma^{\nu\rho} + u^\nu \sigma^{\mu\rho})  +\frac{1}{2} \left( 7 u^\mu u^\nu -\Delta^{\mu\nu}\right) \sigma^{\alpha \beta} \sigma_{\alpha\beta} \right\}
\nonumber\\
    & \qquad\quad + \frac{L_{4,2}}{T^2} \left( u^\mu u^\nu -\frac{\Delta^{\mu\nu}}{3} \right) \sigma^{\alpha \beta} \sigma_{\alpha\beta} \bigg] +\order{\partial^4}\,.
 \label{Int:5}
\end{align}
This above expression precisely cancels the terms in Eq.~\eqref{Int_unstared} up to $\order{\partial^3}$, i.e.,
\begin{equation}
    \int \mathrm{dP}\, p^\nu \mathfrak{C}[f] = (\textit{I} + \textit{II} + \textit{III}) + \textit{IV} + \textit{V} =  \order{\partial^4} \,,
\end{equation}
therefore ensuring the energy-momentum conservation up to third order. 
\end{widetext}

%=======================================
\bibliography{ref}
\end{document}